\documentclass[prl,twocolumn,letterpaper]{revtex4}

\usepackage{graphicx}
\usepackage{tabularx}
\usepackage{amsmath}

\newcommand{\ion}[2]{\mbox{$^{#2}$#1$^+$}}

\newcommand{\Yb}[1]{\ion{Yb}{#1}}

\newcommand{\lev}[3]{\mbox{#2$_{\mbox{\tiny$#3$}}$}}
\newcommand{\levs}{\lev{2}{S}{1/2}}
\newcommand{\levp}{\lev{2}{P}{1/2}}
\newcommand{\levd}{\lev{2}{D}{3/2}}

\newcommand{\unit}[1]{\,\mbox{#1}}
\newcommand{\Hz}{\unit{Hz}}
\newcommand{\kHz}{\unit{kHz}}
\newcommand{\MHz}{\unit{MHz}}
\newcommand{\GHz}{\unit{GHz}}

\newcommand{\uW}{\unit{$\mu$W}}

\newcommand{\cm}{\unit{cm}}

\newcommand{\um}{\unit{$\mu$m}}
\newcommand{\nm}{\unit{nm}}

\newcommand{\s}{\unit{s}}
\newcommand{\ms}{\unit{ms}}
\newcommand{\us}{\unit{$\mu$s}}

\newcommand{\G}{\unit{G}}

\newcommand{\etal}{{\em et al.}}

\newcommand{\ish}{\mbox{$\sim$}\,}

\newcommand{\lr}{\mbox{$\leftrightarrow$}}
\newcommand{\bra}[1]{\mbox{$\left< #1 \right|$}}
\newcommand{\ket}[1]{\mbox{$\left| #1 \right>$}}

\newcommand{\beq}{\begin{equation}}
\newcommand{\eeq}{\end{equation}}

\begin{document}

\title{Simple Manipulation of a  Microwave Dressed-State Ion Qubit}

\author{S. C. Webster}
\author{S. Weidt}
\author{K. Lake}
\author{J. J. McLoughlin}
\author{W. K. Hensinger}
\affiliation{Department of Physics and Astronomy, University of Sussex, Brighton, BN1 9QH, UK}

\date{\today}

\begin{abstract}
Many schemes for implementing quantum information processing require that the atomic states used have a non-zero magnetic moment, however such magnetically sensitive states of an atom are vulnerable to decoherence due to fluctuating magnetic fields. Dressing an atom with an external field is a powerful method of reducing such decoherence [N. Timoney \etal, Nature {\bf 476}, 185], even if the states being dressed are strongly coupled to the environment. We introduce an experimentally simpler method of manipulating such a dressed-state qubit, which allows the implementation of general rotations of the qubit, and demonstrate this method using a trapped ytterbium ion. 
\end{abstract}


\maketitle

A key component of any quantum information processor (QIP) is a long-lived qubit, isolated from the environment to protect against decoherence \cite{00:divincenzo}. For QIP devices based on trapped ions the dominant source of decoherence is usually dephasing due to random fluctuations of the magnetic field surrounding the ions unless a `clock' qubit, formed from a pair of levels whose energy separation is unaffected by a change in magnetic field, is used. In many circumstances however, such clock qubits cannot be used, either because they do not exist in the ion in question, or they are not compatible with the process by which multiple-ion quantum gates are to be performed. The static magnetic field gradient gate proposed by Mintert \etal~\cite{01:mintert} is a promising gate technology from the point of scaling up an ion based quantum computer. A static magnetic field gradient is applied to the ions, which serves two purposes; it allows individual addressability of ions by a global microwave field \cite{09:johanning}, and it allows microwave fields to produce changes to the external degrees of freedom of the ions, mediating gate operations. The different states making up the qubit however are required to have different magnetic moments, meaning a clock qubit cannot be used. Gates using this static field gradient method have been implemented, however the gate fidelity was limited by field-induced decoherence \cite{12:khromova}.

Although the intrinsic coherence time of a qubit may be short, methods exist to increase it, such as dynamical decoupling by applying a series of $\pi$ pulses to the qubit \cite{99:viola, 07:uhrig}. Recently, Timoney \etal~\cite{11:timoney} proposed and implemented a scheme where two states with opposite magnetic moments are dressed by continuous microwave fields to form a state whose energy has no field dependence; this state is then combined with a state which intrinsically has no field dependence to form an effective clock qubit. A method of manipulating the dressed-state qubit was also described, which allows rotation around a specific axis in the $x-y$ plane of the Bloch sphere only. Timoney \etal~\cite{11:timoney} reported a two orders of magnitude improvement of coherence time using this dressed-state qubit, which should allow high-fidelity multi-qubit gates based on static field gradients to be performed.

\begin{figure}[tbhp]
\begin{center}
\includegraphics[width=0.45\textwidth]{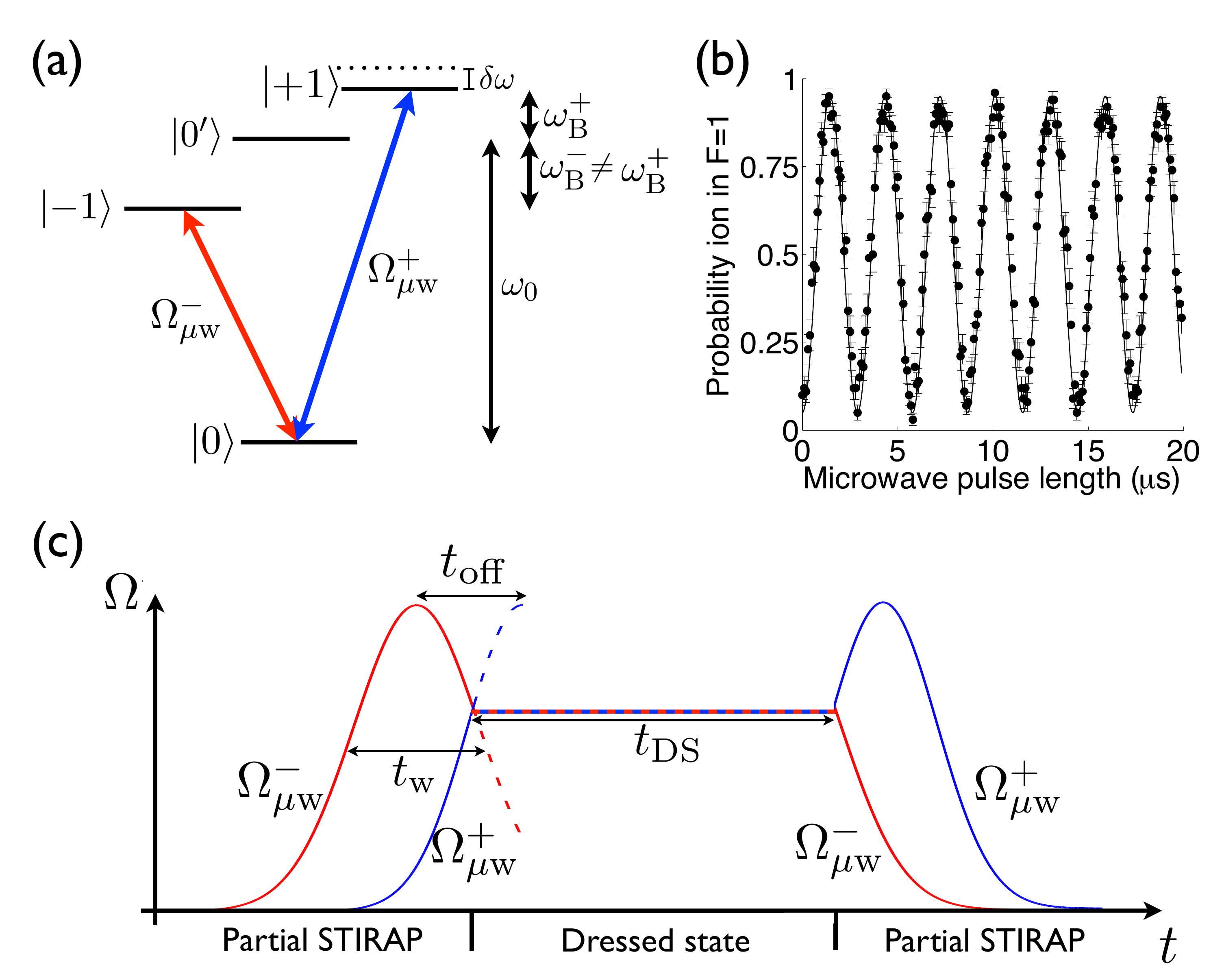}
\caption{
(a) The \levs~ground state of the \Yb{171} ion consisting of the $F=0$ state \ket{0} and three $F=1$ states \ket{-1}, \ket{0'} and \ket{+1} whose degeneracy is lifted by an applied magnetic field. The hyperfine splitting $\omega_0/2\pi$ is $12.6\GHz$ and $\omega_{\rm B}^+/2\pi$ is $13.7\MHz$ for a field of 9.8\G. Due to the 2nd order Zeeman shift $(\omega_{\rm B}^+-\omega_{\rm B}^-)/2\pi$ is $-30\kHz$. Resonant microwave fields can be applied to manipulate or dress the ion, and a radiofrequency field can drive transitions between the $F=1$ levels.
(b) Rabi oscillations between the first-order magnetic field insensitive \ket{0} and \ket{0'} states, with a Rabi frequency of $2\pi\times342\kHz$.
(c) Schematic of the dressed-state pulse sequence. A partial STIRAP process transfers the ion from \ket{+1} to \ket{D}. The microwave Rabi frequencies are held constant for a time $t_{\rm DS}$, during which time manipulation of the dressed-state qubit can take place using the rf field. Finally the STIRAP process is completed, transferring any population in the \ket{D} state to \ket{-1}. The STIRAP pulses are Gaussian, of width $t_{\rm w}$ and offset $t_{\rm off}$.
}
\label{fig:4level}
\end{center}
\end{figure}

We describe a different method of manipulating the dressed-state qubit which allows direct rotation of the qubit state about any axis in the $x-y$ plane. In addition to allowing more general rotations, this method is simpler to implement experimentally, removing requirements on the setting of the initial relative phases of the driving fields that are needed in the original manipulation method \cite{11:timoney}. We demonstrate an experimental implementation of the dressed-state qubit, and our manipulation scheme, using a single \Yb{171} ion.

To describe our modification to the idea first proposed by Timoney \etal~\cite{11:timoney}, we start first by summarising the method they have proposed, in order to highlight the differences.

The creation and manipulation of the dressed-state qubit is presented here as applied to the ground state of \Yb{171}, although the method is applicable to other systems. The hyperfine structure of the \levs~ground state of \Yb{171} is shown in figure \ref{fig:4level}(a) and consists of an $F=0$ level \ket{0}, and three levels with $F=1$ (\ket{-1}, \ket{0'} \& \ket{+1}, labeled corresponding to their $m_{\rm F}$ value). The degeneracy of these levels is lifted by a static magnetic field. There are two classes of transitions within the system, microwave transitions between $F=0$ and the different $F=1$ levels, and radiofrequency (rf) transitions between the $F=1$ levels. 

By applying continuous microwave excitation to dress the ion, magnetically sensitive states can be stabilised against disturbance caused by magnetic field fluctuations. The three atomic states \ket{0},  \ket{-1} and \ket{+1} are dressed by two microwave fields of equal Rabi frequency $\Omega_{\rm \mu w}$ resonant with the \ket{0}\lr\ket{-1} and \ket{0}\lr\ket{+1} transitions, resulting in a Hamiltonian $H_{\rm \mu w} = \frac{\hbar\Omega_{\rm \mu w}}{2}(\ket{+1}\bra{0}+\ket{-1}\bra{0} + {\rm h.c.})$, setting the two microwave phases set to zero (all Hamiltonians are presented in the interaction picture and after making the rotating wave approximation). The eigenstates of the coupled system are \cite{11:timoney}
\begin{align}
\ket{D} & = \frac{1}{\sqrt{2}}(\ket{+1} - \ket{-1})\\
\ket{u} & = \frac{1}{2}\ket{+1} + \frac{1}{2}\ket{-1} + \frac{1}{\sqrt{2}}\ket{0} \\
\ket{d} & = \frac{1}{2}\ket{+1} + \frac{1}{2}\ket{-1} - \frac{1}{\sqrt{2}}\ket{0} 
\end{align}
and the Hamiltonian can be re-written in this basis as
\begin{equation}
H_{\rm \mu w} = \frac{\hbar\Omega_{\rm \mu w}}{\sqrt{2}} (\ket{u}\bra{u} - \ket{d}\bra{d}). \label{eq:dressedHam}
\end{equation}

Without the dressing fields, fluctuations of the magnetic field would cause the \ket{D} superposition to precess to the state $(\ket{-1}+\ket{+1})/\sqrt{2}=(\ket{u}+\ket{d})/\sqrt{2}$, however the dressing microwaves lift the degeneracy of \ket{D}, \ket{u} and \ket{d} so only the part of the fluctuation spectrum around this splitting frequency of $\Omega_{\rm \mu w}/\sqrt{2}$ will cause the ion to leave the state \ket{D}. The dressing fields thus protect the ostensibly field sensitive state \ket{D} from field fluctuations.

The remaining state \ket{0'} does not couple to this dressed subsystem without additional interactions. If the states \ket{0'} and \ket{D} are used as qubit states then the qubit phase will be unaffected by magnetic field fluctuations (besides those bridging the energy gap).

Timoney \etal~\cite{11:timoney} described a method to manipulate the dressed-state qubit as follows.  To first order in the applied magnetic field, the transition frequencies linking \ket{-1}, \ket{0'} and \ket{+1},  $\omega_{B}^-$ and $\omega_{B}^+$, are equal, so a single rf field of Rabi frequency $\Omega_{\rm rf}$ and phase $\phi_{\rm rf}$ will couple all three of the $F=1$ states. The resultant Hamiltonian $H=H_{\rm \mu w}+H_{\rm rf}$ where the rf terms to add to the microwave terms (\ref{eq:dressedHam}) are:
\begin{align}
H_{\rm rf} & = \frac{\hbar\Omega_{\rm rf}}{2}(e^{i\phi_{\rm rf}}\ket{-1}\bra{0'}+e^{-i\phi_{\rm rf}}\ket{+1}\bra{'0} + {\rm h.c.}) \label{eq:rfHam} \\
  & = \frac{\hbar\Omega_{\rm rf}}{2} ( \cos{\phi_{\rm rf}}(\ket{u} + \ket{d})\bra{0'} - \sqrt{2}i\sin{\phi_{\rm rf}}\ket{D}\bra{0'} + {\rm h.c.})
\end{align}
after rewriting in the dressed-state basis.

The states \ket{-1} and \ket{+1} were already linked together by the microwave fields, so adding a second linkage between them by applying an rf field results in a looped system. This loop means that the resulting form of the interaction between states is drastically affected by the phase of the rf, the different paths around the loop interfering. This interference means the phase $\phi_{\rm rf}$ controls the Rabi frequency at which a specific rotation in the Bloch sphere occurs (and also the Rabi frequency at which \ket{0'} is off-resonantly coupled out of the qubit subspace, to the states \ket{u} and \ket{d}). Setting the phase $\phi_{\rm rf}$ to $\pi/2$ produces maximum coupling between the two qubit states and no coupling out of the qubit manifold, resulting in an interaction $H_{\rm rf}=\frac{\hbar\Omega_{\rm rf}}{\sqrt{2}}i(\ket{0'}\bra{D}-\ket{D}\bra{0'})$. This is a $\sigma_y$ coupling, rotating the qubit about the $y$ axis of the Bloch sphere. When implemented experimentally, care must be taken whenever the rf or microwave frequencies are changed that $\phi_{\rm rf}$ is correctly set to $\pi/2$ to produce the desired rotation.

In contrast, when a normal two-level system is resonantly driven, the phase of the driving field controls the axis in the $x-y$ plane of the Bloch sphere about which the state rotates (a $\sigma_\phi=\cos{\phi}\,\sigma_x+\sin{\phi}\,\sigma_y$ coupling from hereon; $\sigma_x$ and $\sigma_y$ being specific cases of this more general coupling). This gives a flexibility in qubit rotation that is not present in Timoney \etal's method \cite{11:timoney}.

Timoney's method \cite{11:timoney} as presented assumes that $\hbar\omega_{\rm B}^+$, the separation in energy between \ket{0'} and \ket{+1},  is the same as $\hbar\omega_{\rm B}^-$, the separation between \ket{-1} and \ket{0'}. For a sufficiently large magnetic field however, the 2nd order Zeeman shift lifts this degeneracy to producing a significant difference between the transition frequencies $\delta\omega = \omega_{\rm B}^+ - \omega_{\rm B}^-$.  Unless $\Omega_{\rm rf}\gg |\delta\omega|$ an additional rf field is needed, so both transitions can be resonantly addressed. This doubling of the numbers of rf fields could potentially greatly complicate experiments, for instance in the case where this gate is extended to perform multi-qubit gates \cite{11:timoney}. A two-ion M\o lmer-S\o rensen gate \cite{00:sorensen} would potentially require 8 fields of different frequency due to this effect, rather than 4.

Here we present a simpler method of performing single qubit gate operations. It requires only one rf field and does not require the relative phases of the driving fields to be set to specific values. Arbitrary $\sigma_\phi$ couplings are obtained with a simple change of the rf phase.

This method takes advantage of the non-equal frequencies $\omega^+_B$ and $\omega^-_B$ of the two rf transitions due to the 2nd order Zeeman shift. If we have a single rf field, resonant with \ket{0'} \lr~\ket{+1}, then with the condition that the Rabi frequency $\Omega_{\rm rf}\ll |\delta\omega|$ we can ignore directly driven transitions from \ket{0} to \ket{-1} as off-resonant, changing the rf part of the Hamiltonian (\ref{eq:rfHam}) to
\begin{align}
H_{\rm rf}  = &\, \frac{\hbar\Omega_{\rm rf}}{2}(e^{-i\phi_{\rm rf}}\ket{+1}\bra{0'} + {\rm h.c.}) \\
= &\, \frac{\hbar\Omega'_{\rm rf}}{2}(e^{-i\phi_{\rm rf}}\ket{D}\bra{0'} + {\rm h.c.})\nonumber \\
  & + \frac{\hbar\Omega'_{\rm rf}}{2\sqrt{2}}(e^{-i\phi_{\rm rf}}(\ket{u} + \ket{d})\bra{0'} + {\rm h.c.})
\end{align}
in the dressed-state basis where $\Omega'_{\rm rf}=\Omega_{\rm rf}/\sqrt{2}$.

If $\Omega_{\rm rf} \ll \Omega_{\rm \mu w}$ then transitions from \ket{0} to \ket{d} and \ket{u} are suppressed by the energy gap, and we are left with a resonant interaction between \ket{0} and \ket{D}, with a Rabi frequency $\Omega'_{\rm rf}$. The rf field however now no longer links \ket{-1} to \ket{+1} so there is no loop and $\phi_{\rm rf}$ can be freely chosen without causing interference effects. Changes to $\phi_{\rm rf}$ produce arbitrary $\sigma_\phi$ couplings as occurs in a driven two-level system. Thus, as long as $\Omega_{\rm rf} \ll |\delta\omega|,\Omega_{\rm \mu w}$, a single rf field resonant with \ket{0'} \lr~\ket{+1} is able to manipulate the dressed state qubit, and by changing the phase of the rf perform arbitrary $\sigma_\phi$ rotations. Detuning the rf field would allow general qubit rotations, about any axis in the Bloch sphere.

We demonstrate this manipulation method of the dressed state qubit using a single \Yb{171} ion confined in a linear Paul trap \cite{11:mcloughlin}. Preparation of the ion's initial state and measurement of its final state are performed on the bare ion without the presence of the dressing fields. We will briefly describe these steps and the method used to switch between the bare and dressed states, before describing the dressed-state ion manipulation.

After Doppler cooling, 369\nm~light resonant with \levs, $F=1$ $\rightarrow$ \levp, $F=1$ clears population out of \ket{-1}, \ket{0'} and \ket{+1} and prepares the ion in \ket{0}. Any decay of population into \levd~ is returned to the \levs \lr~\levp~cycle using light at 935\nm.

The state of the ion is measured using a fluorescence technique. This technique allows us to distinguish between the ion being in the $F=0$ state \ket{0} or one of the $F=1$ states \ket{-1}, \ket{0'} and \ket{+1} when using 8\uW~of 369\nm~light focused to a beam waist of 20\um~and resonant with the \levs, $F=1$ $\rightarrow$ \levp, $F=0$ cycling transition (provided the ion is repumped from \levd). If the ion is in one of the $F=1$ states photons will be scattered and detected using a photo-multiplier tube. However, if the ion is in the $F=0$ state the light is $\approx$14.7\GHz~off resonant and no photons will be detected. A simple threshold can then be used to decide if the ion is fluorescing. For our typical parameters three or more photons denotes a bright ion, two or fewer a dark ion. The efficiency of this technique is limited by off resonant excitation which can cause the ion to transition between the fluorescing $F=1$ states and the non-fluorescing $F=0$ state (and vice-versa) \cite{08:myerson}. With our setup we currently achieve a detection fidelity of up to $\approx$ 0.93. This could be improved by increasing the collection efficiency of our imaging optics and reducing the dark count rate; further improvements are also available if the arrival times of collected photons are taken into account in determining the state \cite{06:acton, 08:myerson}.

To drive transitions between \ket{0} and one of the $F=1$ states, a microwave signal is applied to the ion using a microwave horn positioned 4\cm~from the ion's position. 

Figure \ref{fig:4level}(b) shows a typical set of Rabi oscillations with a Rabi frequency of $2\pi\times 342\kHz$~obtained by driving the magnetic field insensitive \ket{0'} \lr~\ket{0} transition with microwaves at 12.6\GHz. In all the experiments reported here, each repeat of the experimental sequence was started at the same phase of the mains cycle.  The decoherence can be quantified by measuring the coherence times of qubits based on these transitions by performing a series of Ramsey split pulse experiments. For the clock qubit \ket{0}-\ket{0'} a coherence time \ish1.6\s~was obtained, while the \ket{0}-\ket{+1} qubit had a coherence time of \ish40\ms, illustrating the way that magnetic field fluctuations dominate the dephasing of field sensitive states.

To determine the two Zeeman splittings in the F=1 manifold, the precise frequencies of the three microwave transitions were measured, and the magnitude of the magnetic field at the ion determined to be 9.80(1)\G. From these frequency measurements we get a frequency difference in the two Zeeman splittings of -29(1)\kHz. For the $F=1$ states the 2nd order Zeeman shift $\delta\omega = \omega_{\rm B}^+ - \omega_{\rm B}^- = -(2g_J\mu_BB/(2I+1))^2/2\hbar^2\omega_0 =  -2\pi\times0.31\kHz/{\rm G}^2$ \cite{80:woodgate}, matching the measured value. 

\begin{figure}[tb]
\begin{center}
\includegraphics[width=0.45\textwidth]{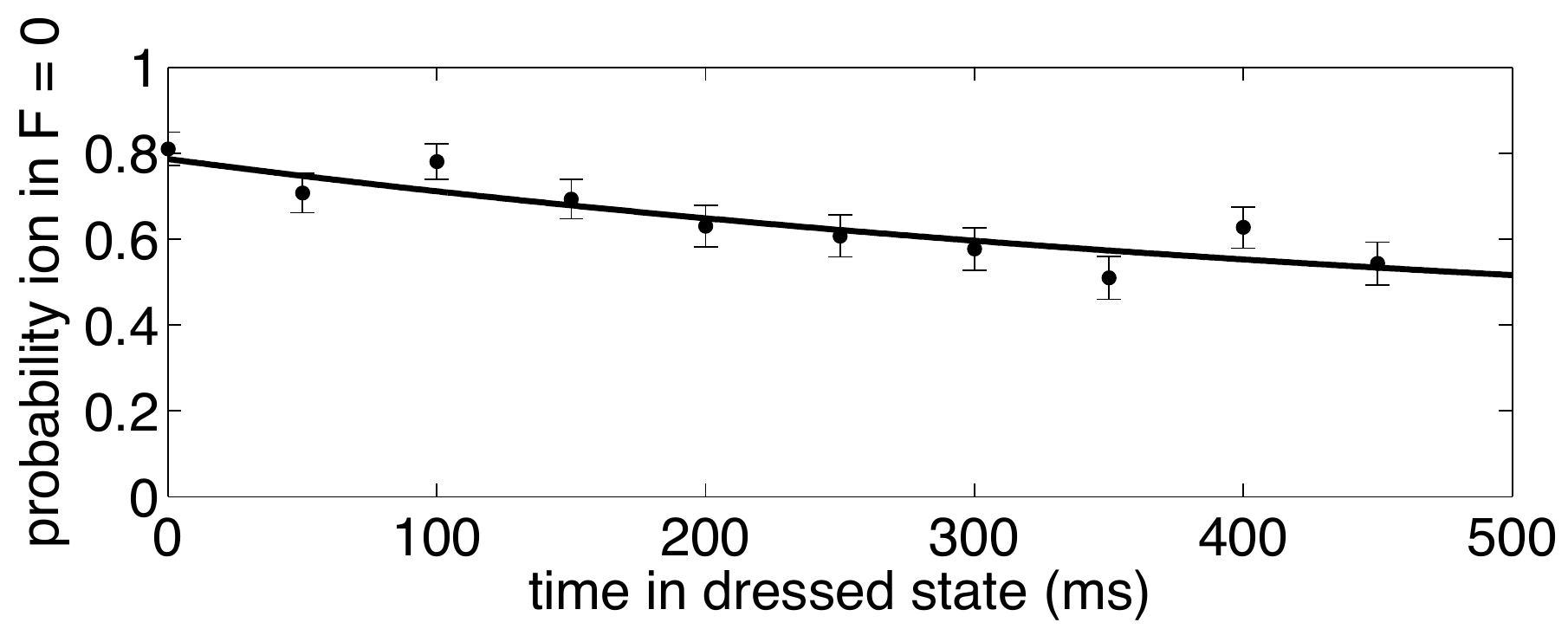}
\caption{
Decay of the \ket{D} state. The ion is held in the \ket{D} state for a variable length of time. After the 2nd partial STIRAP is completed, a $\pi$ pulse swaps the population in states \ket{-1} and \ket{0} before readout. The lifetime of the dressed-state \ket{D} is 550\ms, for microwave Rabi frequencies during $t_{\rm DS}$ of $2\pi\times16\kHz$. The peak microwave Rabi frequency during the STIRAP was $2\pi\times25\kHz$, and the pulses were characterised by $t_{\rm w}=450$\us~and $t_{\rm off}=356$\us.
}
\label{fig:stirap}
\end{center}
\end{figure}

In order to successfully dress the ion we require two microwave dressing fields. To generate the required microwave frequencies, we begin with two different low frequency signals (0-30\MHz) which are then individually amplitude modulated before being combined and then shifted into the microwave domain by mixing with a 12.6\GHz~source before being amplified and sent to the microwave horn. 

An interrupted stimulated Raman adiabatic passage (STIRAP) process \cite{98:bergmann} is used to controllably dress and undress the ion. The ion is prepared in \ket{+1} (using a microwave $\pi$ pulse from \ket{0}) and a STIRAP sequence started by adiabatically modulating the Rabi frequencies $\Omega_{\rm \mu w}^+$ and $\Omega_{\rm \mu w}^-$ of the two microwave fields, as though to transfer the ion to \ket{-1}. At the point at which $\Omega_{\rm \mu w}^+=\Omega_{\rm \mu w}^-$ the ion is in \ket{D}, and the Rabi frequencies are then held constant, `pausing' the STIRAP process. A schematic of this is shown in figure \ref{fig:4level}(c). Once the qubit is to be measured, the Rabi frequency modulation is resumed, finishing the STIRAP to transfer the population in state \ket{D} to \ket{-1} from where it is transferred to \ket{0} by a microwave $\pi$ pulse before the bare state measurement.

The efficiency of the STIRAP process depends on the process being adiabatic. For a peak microwave Rabi frequency of $2\pi\times 25\kHz$, and using Gaussian amplitude envelopes, a maximum transfer efficiency of \ish85\% was obtained for a $t_{\rm w}$ in the range 250-500\us; higher peak Rabi frequencies should improve this transfer efficiency. All data presented here has $t_{\rm w}=450$\us~and $t_{\rm off}=356$\us

A first measure of the robustness of the dressed state qubit is to measure the lifetime of the dressed state \ket{D}. This can be determined by measuring the population of the \ket{D} state as a function of the STIRAP pause time, as shown in figure \ref{fig:stirap}. A fitted exponential to the data points gives a lifetime of \ket{D} of 550\ms, with microwave Rabi frequencies $\Omega_{\rm \mu w}=2\pi\times16\kHz$~dressing the ion. By improving our frequency generation set-up we expect to increase this time to well beyond a second.

\begin{figure}[t]
\begin{center}
\includegraphics[width=0.45\textwidth]{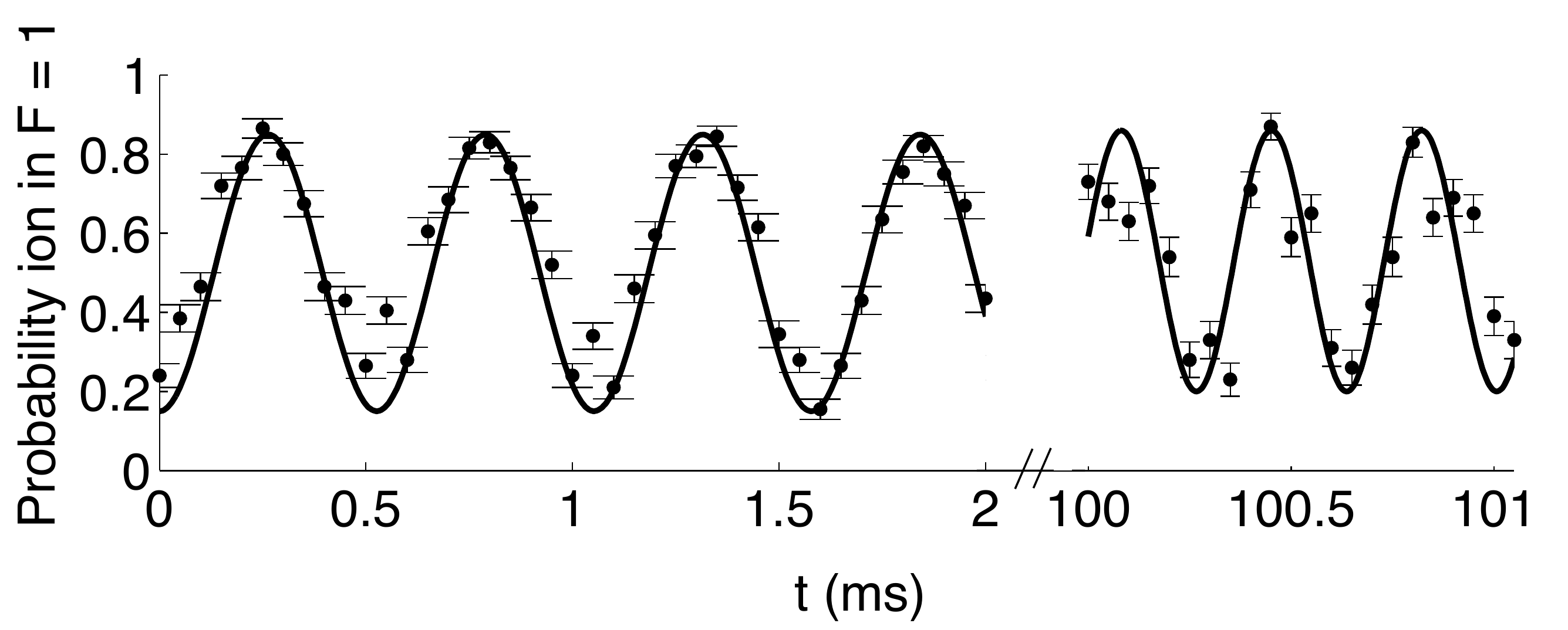}
\caption{
Rabi oscillations within the dressed state.  Based on the short time behaviour, the Rabi frequency $\Omega'_{\rm rf}=2\pi\times1.9\kHz$. The contrast of the Rabi flopping for long pulses is almost unaltered from the short time flopping, indicating lack of dephasing of the qubit. The frequency of oscillation appears changed for pulse lengths over 100\ms~ due to slow fluctuations of experimental parameters while data was being taken.}
\label{fig:DSRabi}
\end{center}
\end{figure}

To coherently manipulate our dressed qubit an rf field tuned on resonance with either  \ket{0'} \lr~\ket{+1} or \ket{0'} \lr~\ket{-1} is required. The rf field is generated by sending a 5$\,{\rm V}_{\rm amp}$ signal into a 3 turn coil positioned \ish1\cm~from the ion. A simple RC network is used to impedance match the circuit and produce a broad resonance around the radio frequencies used.

Figure \ref{fig:DSRabi} shows Rabi oscillations within the dressed state, for both short and long pulses of the rf field, with a Rabi frequency $\Omega'_{\rm rf}=2\pi\times1.9\kHz$. At long flopping times, there is little dephasing of the qubit, however the data departs from a perfect sinusoid due to long timescale fluctuations of the experimental parameters.

\begin{figure}[t]
\begin{center}
\includegraphics[width=0.35\textwidth]{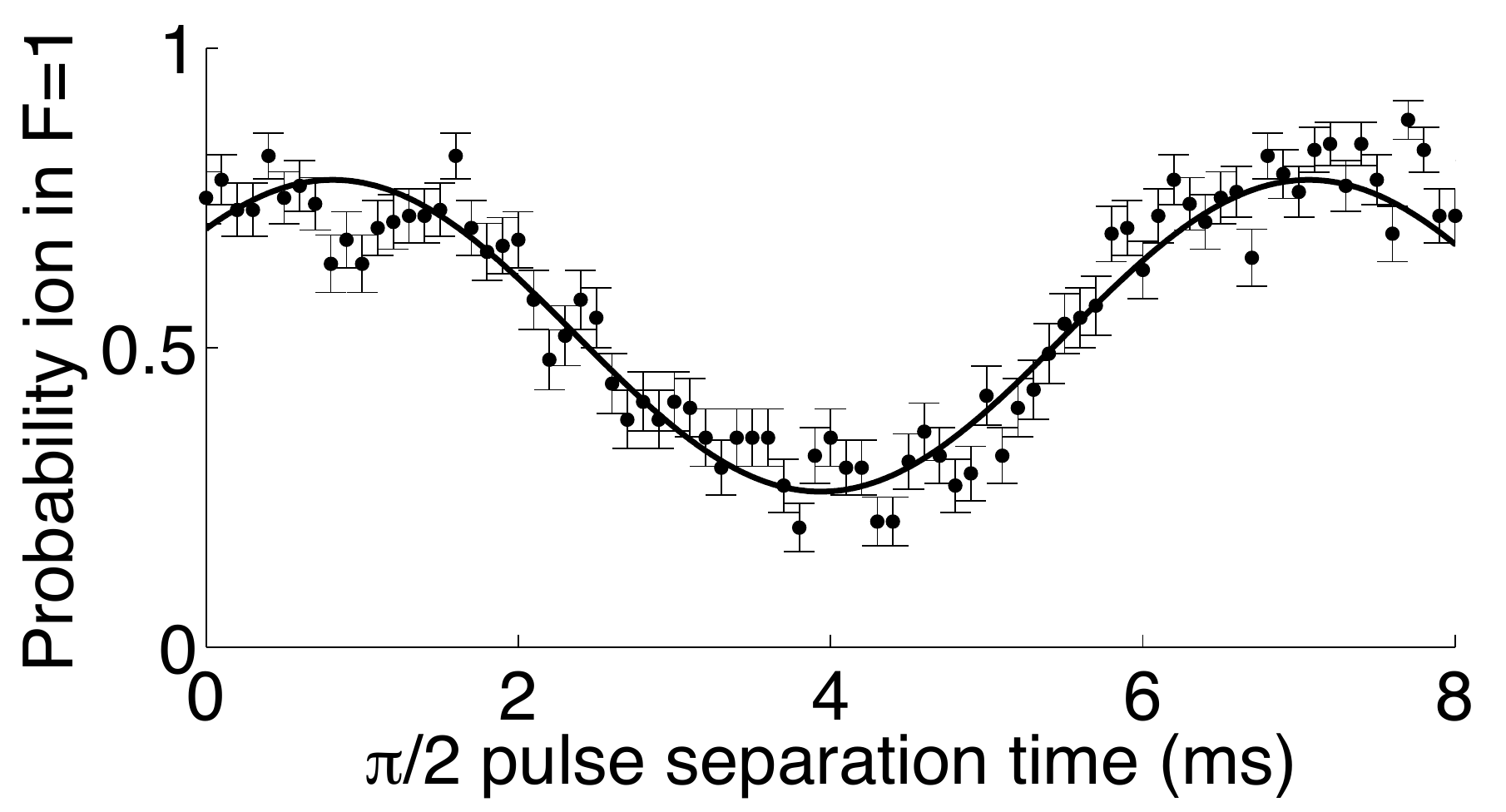}
\caption{
Ramsey fringe within the dressed state. Two $\pi/2$ pulses, detuned from resonance, are separated by a variable time. During the separation time the ion and rf develop a phase difference, causing the second $\pi/2$ pulse to be performed about a different axis in the $x-y$ plane of the Bloch-sphere (a $\sigma_\phi$ rotation). From the fringe period, a detuning $\delta_{\rm rf}=2\pi\times160\Hz$ is inferred.}
\label{fig:DSRamsey}
\end{center}
\end{figure}

To demonstrate the ability to perform arbitrary rotations, we perform a Ramsey split pulse experiment. By detuning the rf field from resonance by $\delta_{\rm rf}\ll\Omega'_{\rm rf}$ the $\pi/2$ nature of the pulses are unaffected, however the ion and the rf develop a relative phase proportional to the time between Ramsey pulses, changing the rotation axis in the Bloch sphere about which the second $\pi/2$ pulse operates and producing a Ramsey fringe as shown in figure \ref{fig:DSRamsey}. 

Microwave dressing the magnetically sensitive levels of an atom allows the construction of a  dressed-state qubit that is robust against decoherence due to magnetic field fluctuations. We have described and implemented a simple single-qubit gate operating on such a dressed-state qubit. It is experimentally simple to implement, requiring only a single rf field to be applied to the ion, and does not require knowledge of the absolute phase of the rf. Arbitrary $\sigma_\phi$ couplings are simply implemented with a change of the phase of the rf field. Using a similar extension to the coupling method as proposed by Timoney \etal~\cite{11:timoney}, our method could be extended to drive multi-ion entangling gates.

This work is supported by the UK Engineering and Physical Sciences Research Council (EP/E011136/1, EP/G007276/1), the European Commission's Seventh Framework Programme (FP7/2007-2013) under grant agreement no. 270843 (iQIT), the Army Research Laboratory under Cooperative Agreement Number W911NF-12-2-0072 and the University of Sussex. The views and conclusions contained in this document are those of the authors and should not be interpreted as representing the official policies, either expressed or implied, of the Army Research Laboratory or the U.S. Government. The U.S. Government is authorized to reproduce and distribute reprints for Government purposes notwithstanding any copyright notation herein.

\bibliography{paperrefs}

\end{document}